\documentclass[twocolumn]{emulateapj}

\usepackage{graphicx}
\usepackage{epsfig}

\newcommand\be{\begin{equation}}
\newcommand\ba{\begin{eqnarray}}
\newcommand\ee{\end{equation}}
\newcommand\ea{\end{eqnarray}}
\newcommand{\midtilde}{\raisebox{-0.25\baselineskip}{\textasciitilde}}

\shorttitle{TESTING THE NO-HAIR THEOREM. IV. IRON LINES}
\shortauthors{JOHANNSEN \& PSALTIS}

\begin{document}

\title{TESTING THE NO-HAIR THEOREM WITH OBSERVATIONS IN THE ELECTROMAGNETIC SPECTRUM.\\ IV. RELATIVISTICALLY BROADENED IRON LINES}

\author{Tim Johannsen,$^{1,2,3,4}$\footnote{CITA National Fellow} and Dimitrios Psaltis,$^{4,5}$}
\affil{$^1$Department of Physics and Astronomy, University of Waterloo, 200 University Avenue West, Waterloo, ON N2L 3G1, Canada; \\
$^2$Canadian Institute for Theoretical Astrophysics, University of Toronto, Toronto, ON M5S 3H8, Canada; \\
$^3$Perimeter Institute for Theoretical Physics, 31 Caroline Street North, Waterloo, ON N2L 2Y5, Canada; \\
$^4$Physics and Astronomy Departments, University of Arizona, 1118 East 4th Street, Tucson, AZ 85721, USA; \\
$^5$Harvard-Smithsonian Center for Astrophysics, 60 Garden Street, Cambridge, MA 02138, USA}

\begin{abstract}

According to the no-hair theorem, astrophysical black holes are fully characterized by their masses and spins and are described by the Kerr metric. This theorem can be tested observationally by measuring (at least) three different multipole moments of the spacetimes of black holes. In this paper, we calculate the profiles of fluorescent iron lines emitted from the accretion flows around black hole candidates within a framework that allows us to perform the calculation as a function of its mass and spin as well as of a free parameter that measures potential deviations from the Kerr metric. We show that such deviations lead to line profiles that are significantly altered and may exhibit a modified flux ratio of the two peaks in their characteristic double-peaked shape. We also show that the disk inclination can be measured independently of the spin and the deviation parameter at low to intermediate inclination angles as in the case of Kerr black holes. We estimate the precision that near-future X-ray missions such as {\em Astro-H} and {\em ATHENA}$+$ are required to achieve in order to resolve deviations from the Kerr metric in iron line profiles and show that constraints on such deviations will be strongest for rapidly spinning black holes. More generally, we show that measuring the line profile with a precision of $\sim 5\%$ at disk inclinations of $30^\circ$ or $60^\circ$ constraints the deviation parameter to order unity for values of the spin $a\gtrsim0.5M$.

\end{abstract}

\keywords{accretion, accretion disks --- black hole physics --- galaxies: active --- gravitation --- line: profiles --- relativistic processes}

\section{INTRODUCTION}

According to the no-hair theorem, black holes are uniquely characterized by their mass $M$ and spin $J$ and are described by the Kerr metric (Israel 1967, 1968; Carter 1971, 1973; Hawking 1972; Robinson 1975). Mass and spin are the first two multipole moments of a black-hole spacetime, and all higher order moments can be expressed in terms of these two (Hansen 1974). As a consequence of the no-hair theorem, all astrophysical black holes are expected to be Kerr black holes.

The defining characteristic of a black hole is its event horizon. Observational evidence for the presence of an event horizon in astrophysical black holes exists (see discussion in, e.g., Psaltis 2006). To date, however, a definite proof of the Kerr nature of these black holes is still lacking. Since mass and spin are the only independent multipole moments of a Kerr black hole, the no-hair theorem can be tested by measuring (at least) three such moments (Ryan 1995).

Several tests of the no-hair theorem have been suggested using observations of gravitational waves from extreme mass-ratio inspirals (Ryan 1995, 1997a,b; Barack \& Cutler 2004, 2007; Collins \& Hughes 2004; Glampedakis \& Babak 2006; Gair et al.\ 2008; Li \& Lovelace 2008; Apostolatos et al.\ 2009; Vigeland \& Hughes 2010; Vigeland et al. 2011; Gair \& Yunes 2011; Rodriguez et al. 2012) and from electromagnetic observations of stars orbiting around Sgr A* (Will 2008; Merritt et al. 2010; Sadeghian \& Will 2011), of pulsar black-hole binaries (Wex \& Kopeikin 1999; Liu et al. 2012), as well as of the quasar OJ287 (Valtonen et al. 2011).

In this series of papers, we develop another class of observational tests of the no-hair theorem in the electromagnetic spectrum, which aim to probe the spacetimes of black holes using the emission from their accretion flows instead of observing companion stars or compact objects at comparatively large distances (see, also, Bambi \& Barausse 2011, Bambi 2011, 2012a,b; Krawczynski 2012). Since we expect deviations from the Kerr spacetime to manifest predominantly in the immediate vicinity of black holes, our approach allows us to make use of the broad range of observables associated with black hole accretion flows.

In the first part of this series (Johannsen \& Psaltis 2010a), we investigated a framework, within which such tests can be carried out. We used a quasi-Kerr metric (Glampedakis \& Babak 2006), which contains an independent quadrupole moment of the form
\begin{equation}
Q=-M\left(a^2+\epsilon M^2\right),
\label{qradmoment}
\end{equation}
where $a\equiv J/M$ and where we have set the gravitational constant $G$ and the speed of light $c$ to unity. In the expression of the quadrupole moment, a potential deviation from the Kerr metric is parameterized in terms of $\epsilon$. In the limit $\epsilon \rightarrow 0$, this metric reduces smoothly to the familiar Kerr metric.

If a measurement yields a nonzero deviation $\epsilon$, then the compact object cannot be a Kerr black hole. Within general relativity as the underlying theory of gravity, it can only be a different type of star, a naked singularity, or an exotic configuration of matter (see Collins \& Hughes 2004; Hughes 2006). If, however, the compact object is otherwise known to be a black hole, i.e., to have a horizon, then a deviation from the Kerr metric implies that both the no-hair theorem and general relativity are violated.

As a first astrophysical application of our framework (Johannsen \& Psaltis 2010b), we simulated images of such objects and demonstrated their dependence on the quadrupole moments of their spacetimes. In particular, we showed that a bright and narrow ring that surrounds the shadow of a black hole with a diameter of about $10M$ (see also, e.g., Beckwith \& Done 2005) has a shape that depends uniquely on the mass, spin, quadrupole moment, and inclination of the black hole (see Psaltis \& Johannsen 2011 and Johannsen 2012 for reviews). Recently, we explored the prospects of imaging this ring using very-long baseline interferometric observations with the {\em Event Horizon Telescope} (Johannsen et al. 2012), a planned global array of (sub-)mm telescopes (Doeleman et al. 2009a,b; Fish et al. 2009). As a second application, we showed how quasi-periodic variability observed in galactic black holes (e.g., Remillard \& McClintock 2006) and active galactic nuclei (AGN; Gierli\'nski et al. 2008) can be used to test the no-hair theorem in two different scenarios (Johannsen \& Psaltis 2011a).

Relativistically broadened fluorescent iron lines that originate from the irradiation off the accretion disks of black holes provide another mechanism to probe the spacetimes of compact objects. Such iron lines have been observed in both galactic black holes and AGN (see Reynolds \& Nowak 2003; Miller 2007 for reviews). Within general relativity, these line profiles can be used to measure the spins of black holes (Fabian et al. 1989; Stella 1990; Laor 1991; Reynolds \& Nowak 2003; Dov${\rm \check{c}}$iak et al. 2004; Beckwith \& Done 2004, 2005; Brenneman \& Reynolds 2006; Reynolds \& Fabian 2008; Dexter \& Agol 2009; Karas \& Sochora 2010; Dauser et al. 2010).

The shape of a particular iron line profile depends on the spin of the black hole, the radial extend of its (geometrically thin) accretion disk, the disk emissivity, and the disk inclination with respect to the line of sight of a distant observer (e.g., Fabian et al. 1989). The key property that allows for measurements of black hole spins is the low-energy tail of a given iron line profile, which depends on the inner disk radius of the accretion disk. This radius, in turn, is assumed to coincide with the innermost stable circular orbit (ISCO) of the spacetime, which depends exclusively on the spin of the black hole in units of its mass. Therefore, the spin of the black hole can be measured directly from the position of the ISCO, even if other important parameters of the black hole, such as its mass or distance, are unknown (e.g., Reynolds \& Nowak 2003; Brenneman \& Reynolds 2006; Reynolds \& Fabian 2008).

In this paper, we analyze the prospects of using observations of fluorescent iron line profiles to test the no-hair theorem. In Psaltis \& Johannsen (2012) we developed a high-performance ray-tracing algorithm and simulated first iron line profiles using the quasi-Kerr metric (Glampedakis \& Babak 2006). We showed that these profiles can depend significantly on potential deviations from the Kerr metric. The quasi-Kerr metric, however, can only be used for moderately spinning black holes with spins $a\lesssim0.4M$ due to the existence of pathologies at radii $r\sim2M$ (Johannsen \& Psaltis 2010a; Johannsen 2013a). Since most of the observed iron lines indicate high values of the spins (see, e.g., Miller 2007), our initial approach cannot be readily applied to comparing models with data.

Recently, we constructed a Kerr-like metric that is regular everywhere outside of the compact object and, thus, suitable for the description of rapidly spinning black hole candidates (Johannsen \& Psaltis 2011b; hereafter JP11). This metric generally harbors a naked singularity; for small perturbations away from the Kerr metric, this metric describes a black hole (Johannsen 2013a). Similar to the quasi-Kerr metric, our metric depends on an additional parameter $\epsilon_3$ and coincides with the Kerr metric if $\epsilon_3=0$. Here, we incorporate our new metric into our algorithm and simulate iron line profiles over the entire range of spins.

We demonstrate that deviations from the Kerr metric lead to shifts of the measured flux primarily at high energies as well as in the low-energy tail of the line profiles. We demonstrate that these changes can be significant and estimate the required precision of future X-ray missions in order to be able to test the no-hair theorem with fluorescent iron lines for disks of different inclinations. In addition, we show that the disk inclination can be measured independently of the values of the spin and the deviation parameter, at least at low to intermediate inclination angles.

We also show, however, that line profiles for different spins and values of the parameter $\epsilon_3$, chosen in such a way that the spacetimes have the ISCO at the same coordinate radius, are practically indistinguishable even for high values of the spin. We argue that this correlation may be reduced in combination with other observables such as the frequencies of quasi-periodic oscillations (QPOs), which depend only indirectly on the location of the ISCO.

\section{TESTING THE NO-HAIR THEOREM WITH FLUORESCENT IRON LINES}

In this section, we simulate profiles of relativistically broadened fluorescent iron lines\footnote{A selection of simulated iron line profiles is available at {\ttfamily http://www.science.uwaterloo.ca/\midtilde tjohanns/}.} from a geometrically thin accretion disk located in the equatorial plane of the compact object that is described by our metric (JP11). We study signatures of potential deviations from the Kerr metric in terms of the free parameter $\epsilon_3$. We summarize the explicit form of this metric as well as some of its properties in the appendix.

The geometry is the same as in Psaltis \& Johannsen (2012) and Johannsen \& Psaltis (2010b). We consider an observer at a large distance $d$ and an inclination angle $\theta_0$ from the rotation axis of the central object. We define an image plane with Cartesian coordinates $(\alpha_0,\beta_0)$ that is centered at $\phi=0$ and is perpendicular to the line of sight of the observer. We trace rays from the image plane to the equatorial plane using a forth-order Runge-Kutta integrator of the geodesic equations.

\begin{figure}[ht]
\begin{center}
\psfig{figure=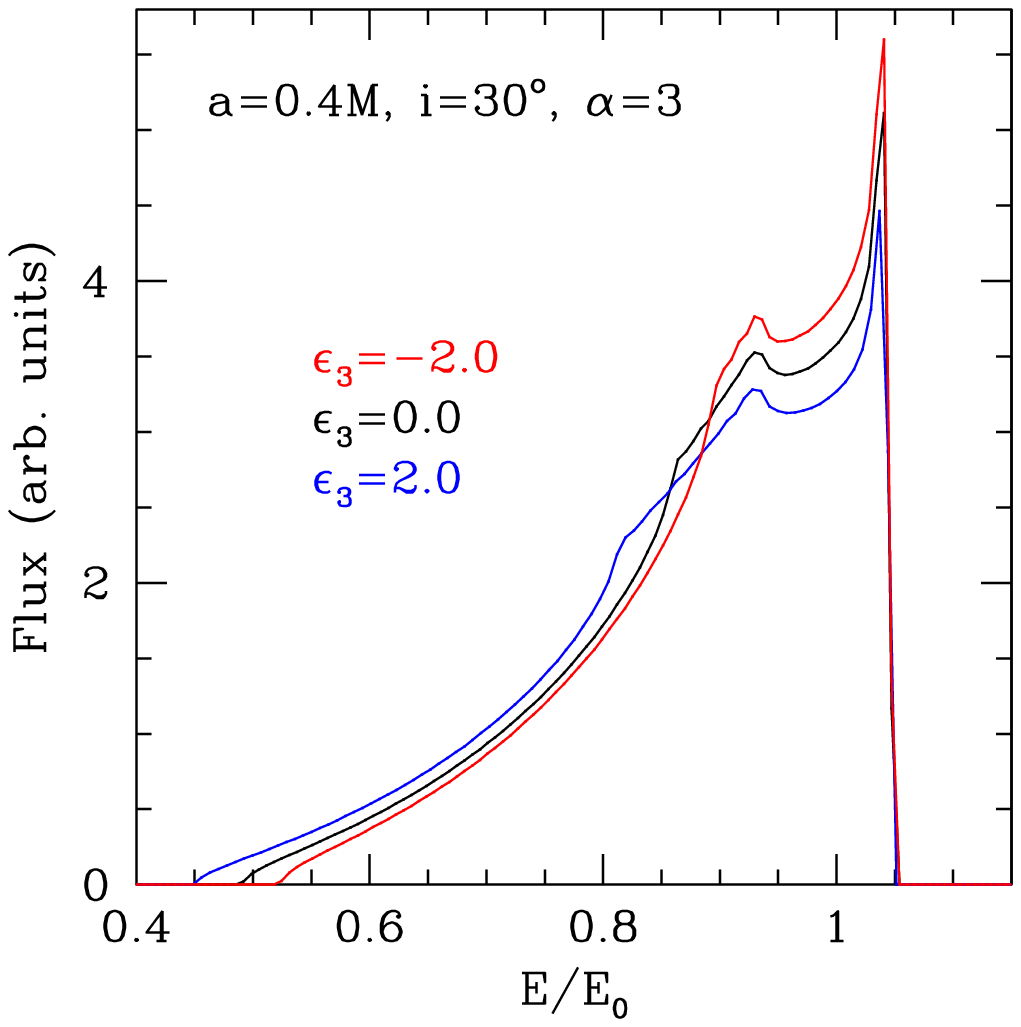,height=3.2in}
\psfig{figure=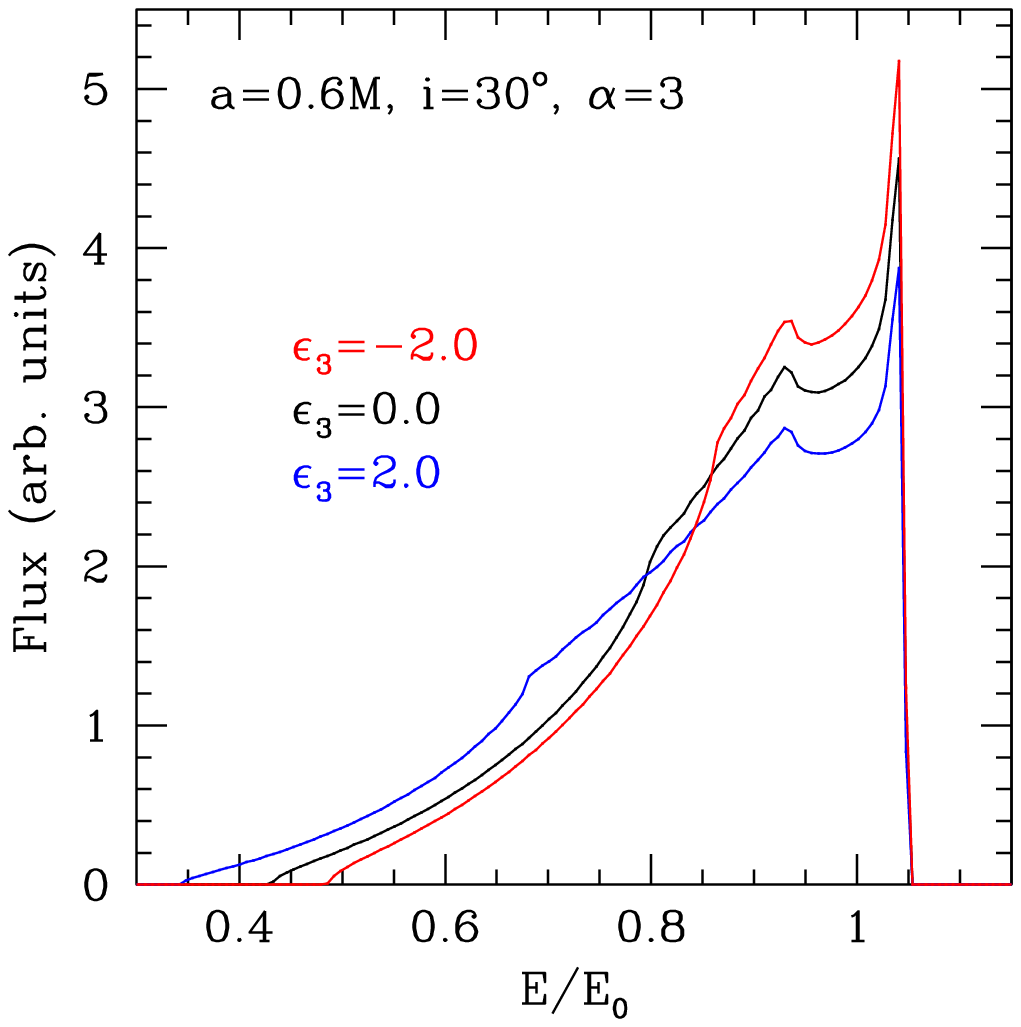,height=3.2in}
\end{center}
\caption{Profiles of fluorescent iron lines for values of the spin (top) $a=0.3M$ and (bottom) $a=0.6M$, inclination $\theta_0=30^\circ$ and emissivity index $\alpha=3$ for several values of the parameter $\epsilon_3$ that measures the deviation from the Kerr metric. Modifications of the profiles occur primarily in the high-energy peak and the low-energy tail.}
\label{lines}
\end{figure}

\begin{figure}[ht]
\begin{center}
\psfig{figure=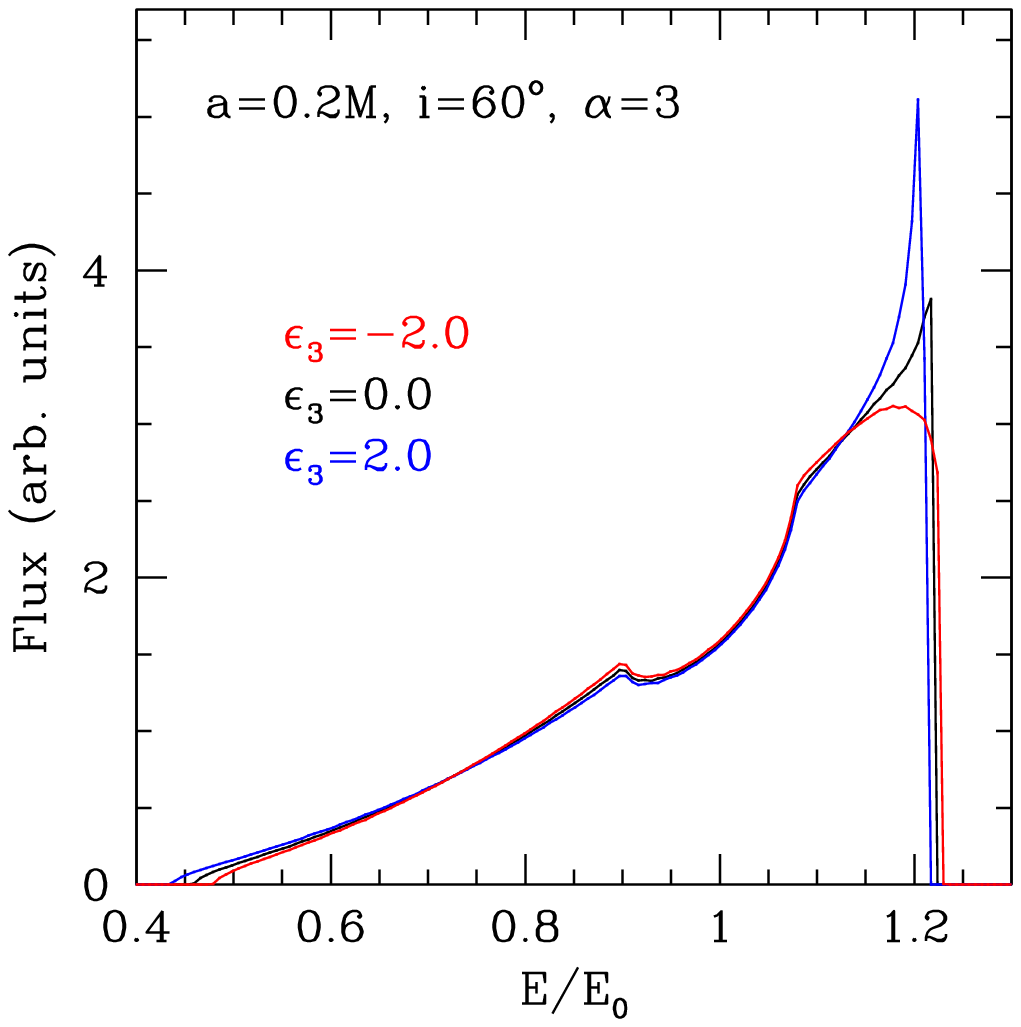,height=3.2in}
\end{center}
\caption{Line profiles for a black hole with a spin $a=0.2M$, inclination $\theta_0=60^\circ$, and emissivity index $\alpha=3$. Modifications of the profile manifest primarily in the high-energy peak.}
\label{bluepeak}
\end{figure}

\begin{figure*}[ht]
\begin{center}
\psfig{figure=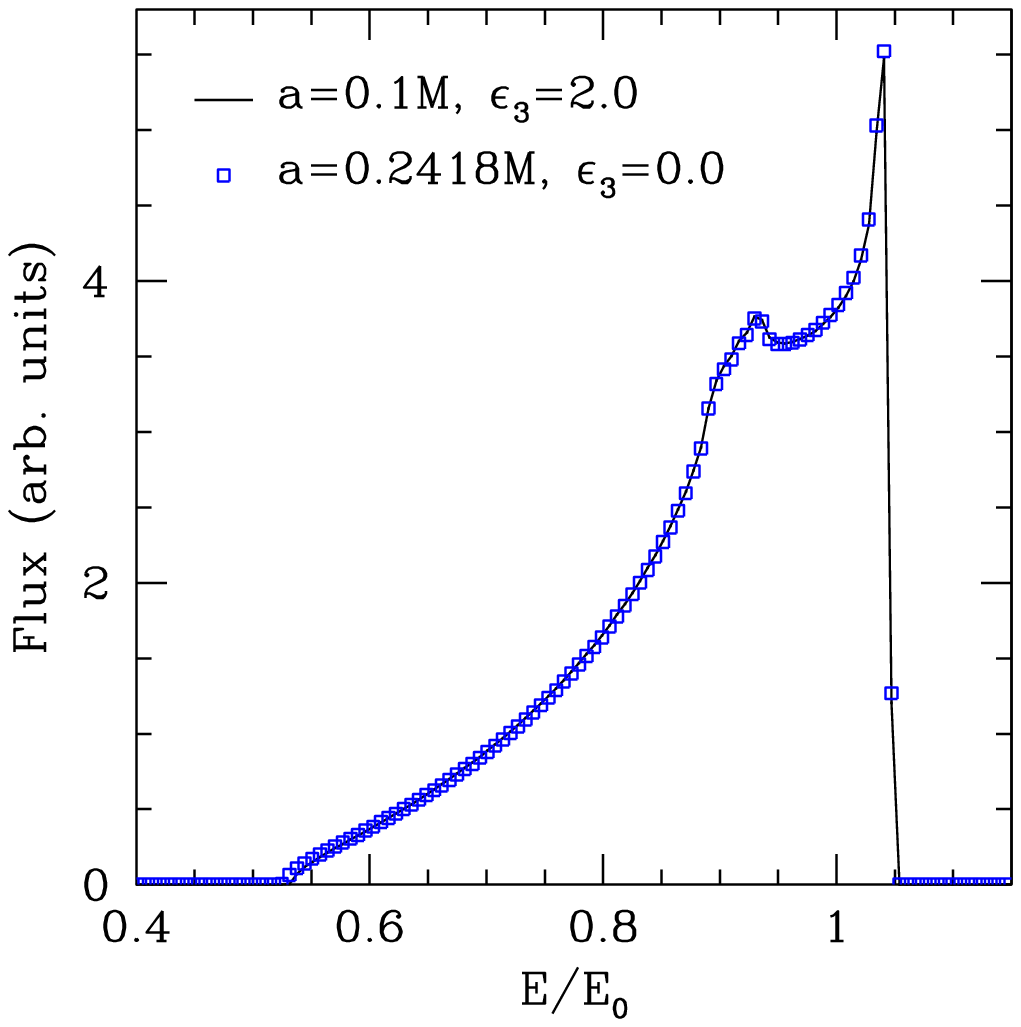,height=2.3in}
\psfig{figure=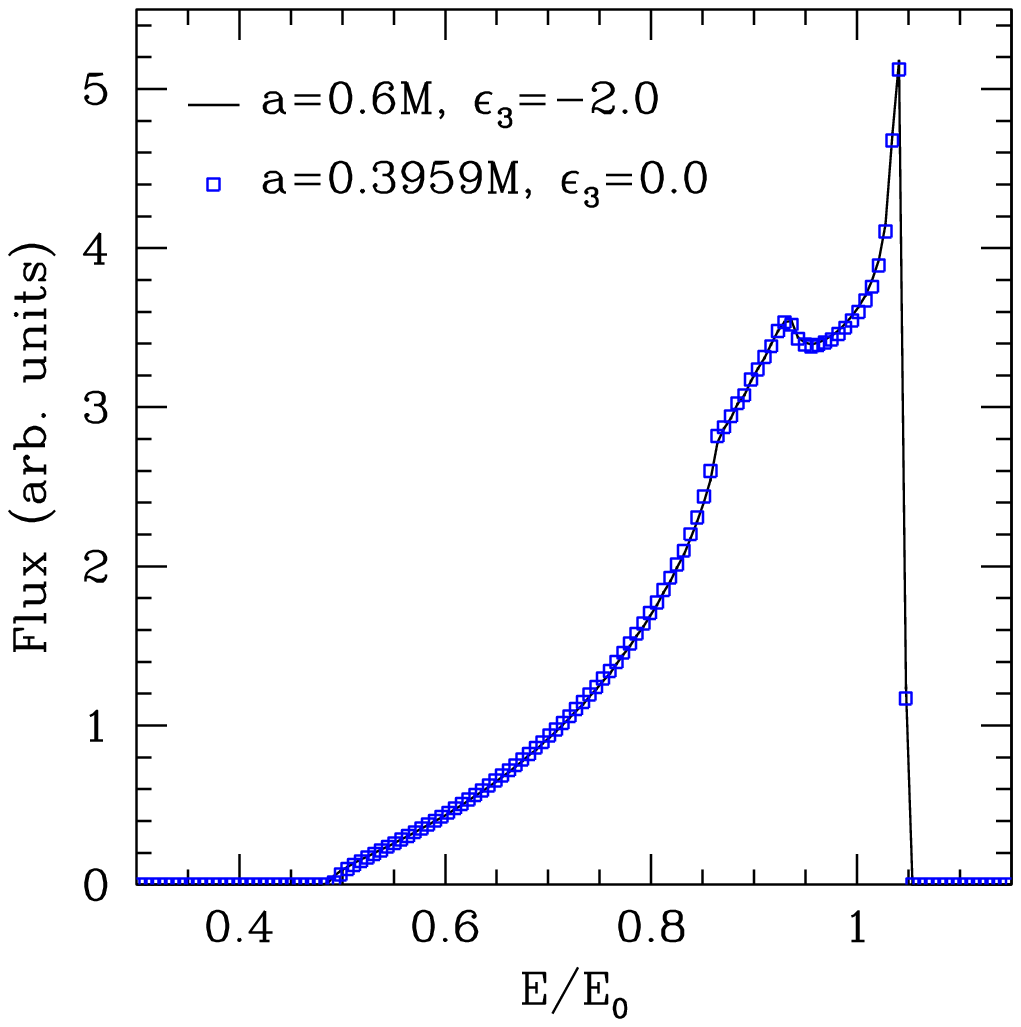,height=2.3in}
\psfig{figure=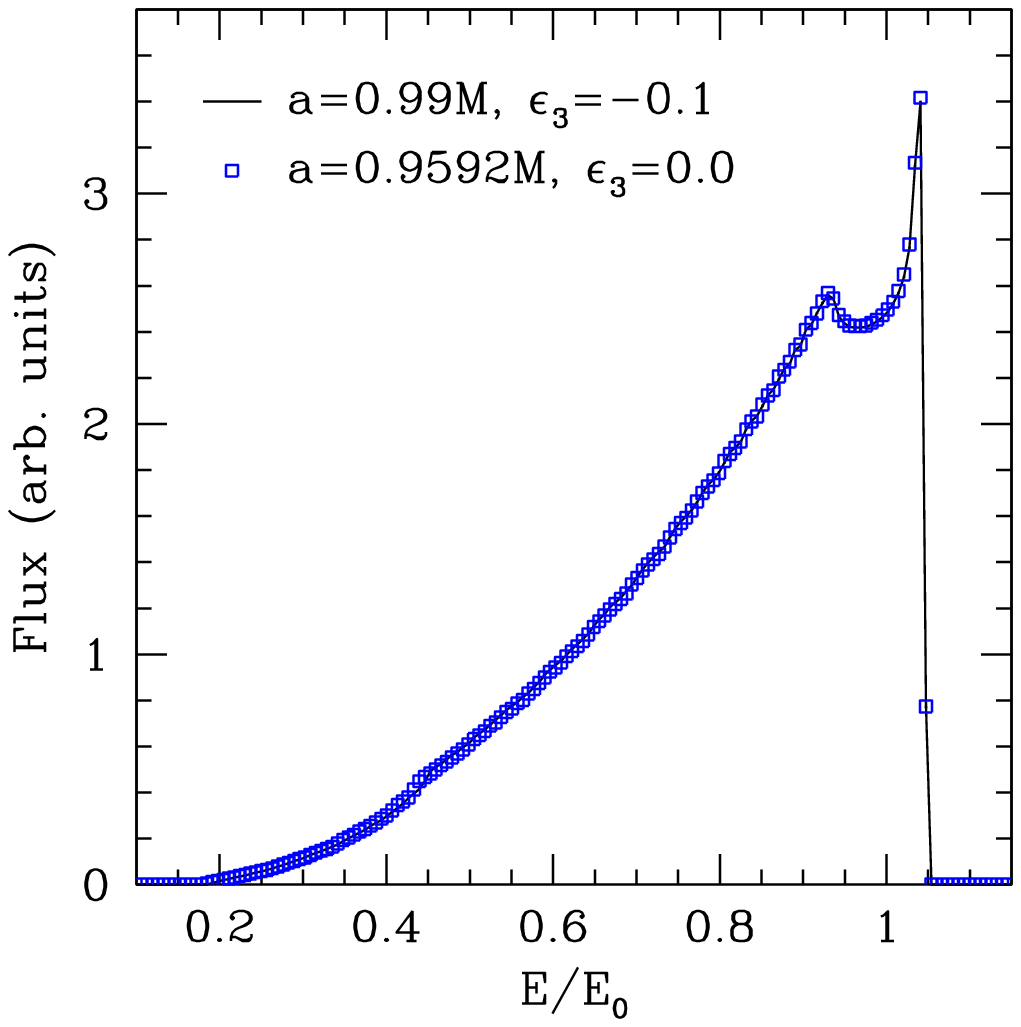,height=2.3in}
\end{center}
\caption{Iron line profiles for several pairs of the spin and values of the parameter $\epsilon_3$ that correspond to the same ISCO radius. The profiles are practically indistinguishable even at a very high spin. In the latter case, however, the range of pairs of the spin and values of the parameter $\epsilon_3$ with the same ISCO is much smaller than at low spins.}
\label{degeneracy}
\end{figure*}

\begin{figure}[ht]
\begin{center}
\psfig{figure=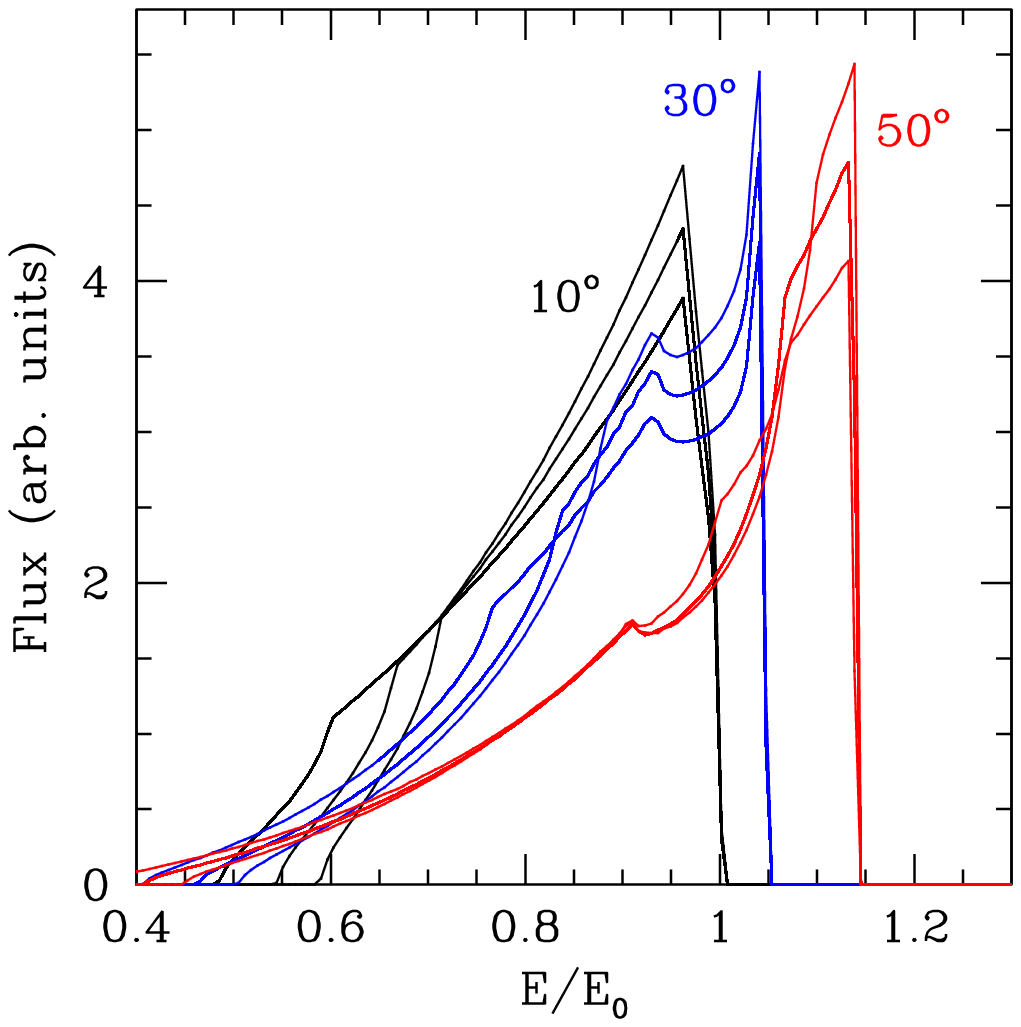,height=3.2in}
\end{center}
\caption{Iron line profiles for a compact object with spin $a=0.5M$ for different values of the observer inclination. For each case, three profiles are shown, corresponding to three different values of the deviation parameter ($\epsilon_3=-2,0,2$). As in the case of Kerr black holes, the extent of the line towards high energies is determined entirely by the inclination and depends very weakly on other system parameters.}
\label{fig:incl}
\end{figure}

We model the accretion disk as having a radial extent from the ISCO to some outer radius $r_{\rm out}$. We calculate the location of the ISCO by numerically finding the minimum of the orbital energy $E$ given by Equation (\ref{energy}). An ISCO exists for most values of the spin and the deviation parameter. For positive values of the spin and sufficiently positive values of the parameter $\epsilon_3$, however, the ISCO disappears, and all circular equatorial orbits are radially stable. Instead, circular equatorial orbits become unstable against small perturbations in the direction vertical to the disk. This part of the parameter space is analyzed in a separate paper (Johannsen 2013b). Given a positive value of the spin, the upper bound on the parameter $\epsilon_3$ up to which an ISCO exists is given in Equation~(\ref{eq:bound}).

At a fixed value of the spin and for positive values of the parameter $\epsilon_3$, the location of the ISCO shifts to smaller radii relative to the location of ISCO in the Kerr metric. For negative values of the parameter $\epsilon_3$, the location of the ISCO shifts to larger radii (JP11). For $\epsilon_3 = 0$, we use the analytic expression for the location of the ISCO in the Kerr metric from Bardeen (1973).

We assume that plasma moves on circular orbits in the disk at the local Keplerian velocity $u = (u^t,0,0,u^\phi)$, where the velocity components $u^t$ and $u^\phi$ are given by the expressions in Equation (\ref{velocity}). We also assume that the disk emission is monochromatic at a rest frame energy $E_0 \approx 6.4~{\rm keV}$ and isotropic with an emissivity \mbox{$\varepsilon(r) \propto r^{-\alpha}$}.

The parameter $\epsilon_3$ likewise affects the lightbending experienced by the emitted photons (Johannsen 2012) as well as the shift of the photon energy due to Doppler boosting, relativistic beaming, and the gravitational redshift, which is given by the expression
\be
g \equiv \frac{E_{\rm im}}{E_{\rm d}} = \frac{ (g_{\mu\nu} k^\mu u^\nu)_{\rm im} }{ (g_{\mu\nu} k^\mu u^\nu)_{\rm d} }.
\ee
In this expression, the subscripts ``im'' and ``d'' refer to the image plane and the accretion disk, respectively, $k^\mu$ is the photon 4-momentum (given explicitly in Johannsen \& Psaltis 2010b), and we set the 3-velocity of the observer at the image plane to zero.

We then compute the observed specific intensity $I(\alpha_0,\beta_0)$ at each point in the image plane using the Lorentz invariant $I/E_p^3$, where $I$ and $E_p$ are the specific intensity and the energy of the emitted photons, respectively. We obtain the total observed specific flux as
\be
F_E = \frac{1}{d^2} \int d\alpha_0 \int d\beta_0 I(\alpha_0,\beta_0) \delta[ E_p - E_0 g(\alpha_0,\beta_0) ].
\ee
We calculate this two-dimensional integral with a Monte Carlo integration over the image plane using a set of narrow energy bins. We simulate the iron line profiles for objects with spin values $0\leq a/M \leq 1$ and values of the parameter $-2 \leq \epsilon_3 \leq 2$ as long as an ISCO exists (see discussion in JP11; Johannsen 2013a).

In Figures~\ref{lines} and \ref{bluepeak}, we plot iron line profiles for several values of the spin and the parameter $\epsilon_3$ for an emissivity index $\alpha=3$ and outer disk radius $r_{\rm out}=100M$ at disk inclinations $\theta_0=30^\circ$ and $\theta_0=60^\circ$, respectively. The effect of changing the parameter $\epsilon_3$ is two-fold. First, the respective fluxes of the ``blue'' peak of the iron line centered at $E/E_0\sim1$ and of the ``red'' peak centered at $E/E_0\sim0.75$ are augmented/diminished for decreasing/increasing values of the parameter $\epsilon_3$. This effect is caused predominantly by the orbital velocity of the accretion flow, which differs from the corresponding orbital velocity in the Kerr metric at the same radius.

At higher values of the inclination, the flux of the ``blue'' peak is predominantly altered, while the flux of the ``red'' peak is affected only marginally. While the absolute flux depends on the particular system, the relative flux between the two peaks is specific to the value of the parameter $\epsilon_3$ and, therefore, a signature of a potential violation of the no-hair theorem. Note, however, that the ``red'' peak is often ill-defined and submerged in the overall profile, which then takes on a more triangular shape. A similar modification of the peak flux can also be achieved by changing the emissivity index $\alpha$ or the outer disk radius $r_{\rm out}$. These parameters have to be determined simultaneously from a spectral fit of the entire line profile.

Second, due to the shift of the location of the ISCO, increasing/decreasing values of the parameter $\epsilon_3$ lead to an extended/shortened ``red tail'' of the iron line at low energies, which is produced by photons emitted at radii near the ISCO that experience the corresponding strong gravitational redshift.

Since the ISCO in our spacetime depends on both the spin and the parameter $\epsilon_3$, a measurement of the ISCO radius alone can only constrain certain combinations of the spin and the parameter $\epsilon_3$. In Psaltis \& Johannsen (2012), we showed that the entire iron line profiles in the quasi-Kerr metric (Glampedakis \& Babak 2006) are practically indistinguishable for different combinations of the spin and the deviation parameter that have the ISCO at the same coordinate radius, but we had to limit our analysis to spin values $a\lesssim0.4M$.

Here, using our new metric, we perform a similar analysis covering low, intermediate, and high spins. In Figure~\ref{degeneracy}, we plot iron line profiles for three such pairs of the spin and the parameter $\epsilon_3$ that have the same ISCO. Even in the case of a very high spin $a=0.99M$ the line profiles are practically indistinguishable. Note, however, that constraints of pairs of spins and values of the parameter $\epsilon_3$ at such high spin values will still be tight, because the range of pairs of spins and values of the deviation parameter that correspond to the same ISCO radius is much smaller than at low spins (see Figure~5 in JP11).

In the case of a Kerr black hole, the location of the ``blue'' edge of the line profile, the sharp drop in observed flux at high energies, depends primarily on the disk inclination. Therefore, the inclination can be measured easily and robustly from the location of the ``blue'' edge, at least for low to moderate values (Fabian et al. 1989). This property also holds for non-zero values of the parameter $\epsilon_3$ as shown in Figure~4.

\section{Precision Requirements for Measurements of Potential Violations of the No-Hair Theorem}

In this section, we estimate the required precision that observations with future X-ray missions have to achieve in order to measure potential deviations from the Kerr metric from fluorescent iron line profiles. We define the relative difference $\sigma$ between a line profile with a nonzero value of the parameter $\epsilon_3$ from the Kerr line profile with otherwise identical parameters by the expression
\be
\sigma \equiv \frac{ \sum_{i=1}^N | F^{\rm dev}_i - F^K_i | }{ \sum_{i=1}^N F^K_i },
\ee
where $N$ is the number of energy bins, and $F^K_i$ and $F^{\rm dev}_i$ are the flux of the $i$th bin of the line in the Kerr metric and the metric JP11, respectively. Here we use \mbox{$N\sim100$}, although the actual value affects the results only marginally.

\begin{figure}[ht]
\begin{center}
\psfig{figure=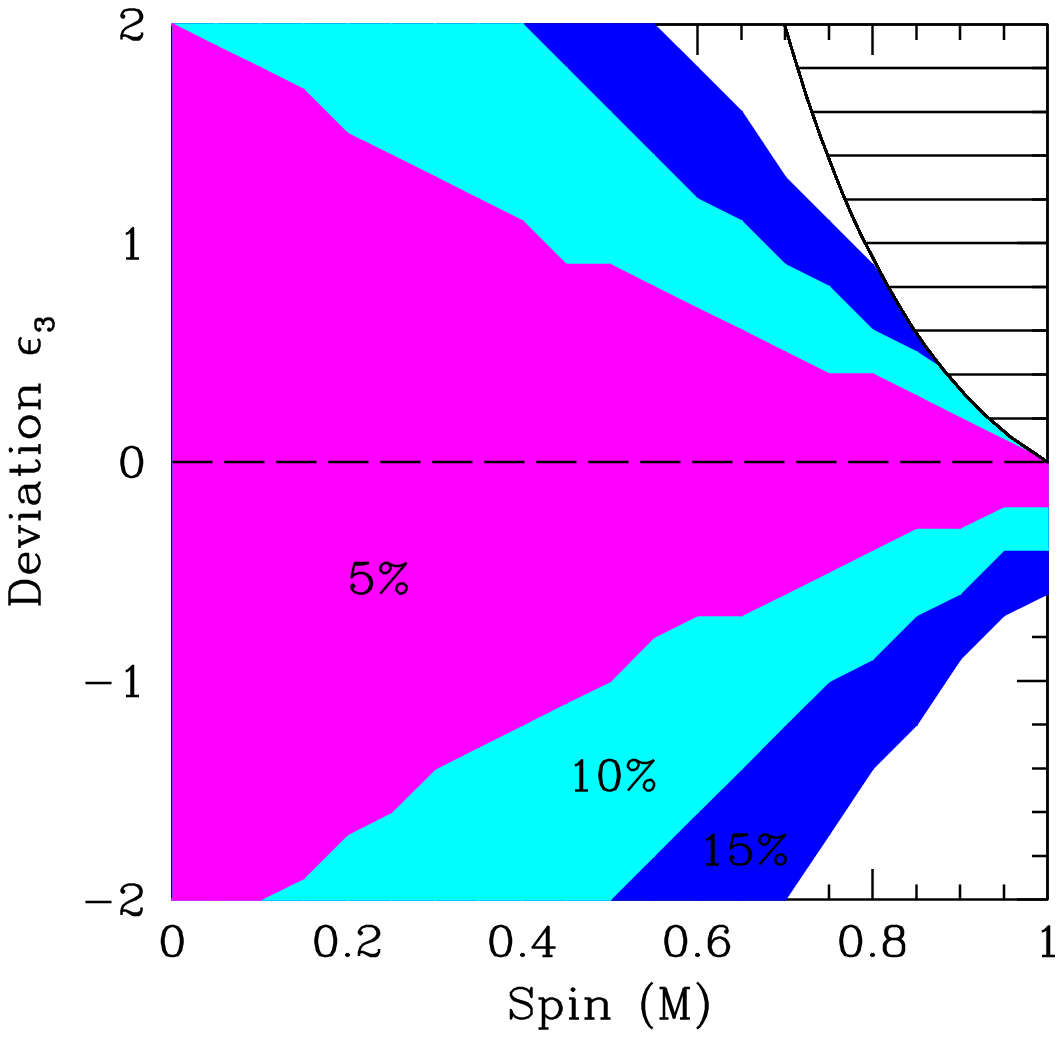,height=3.2in}
\psfig{figure=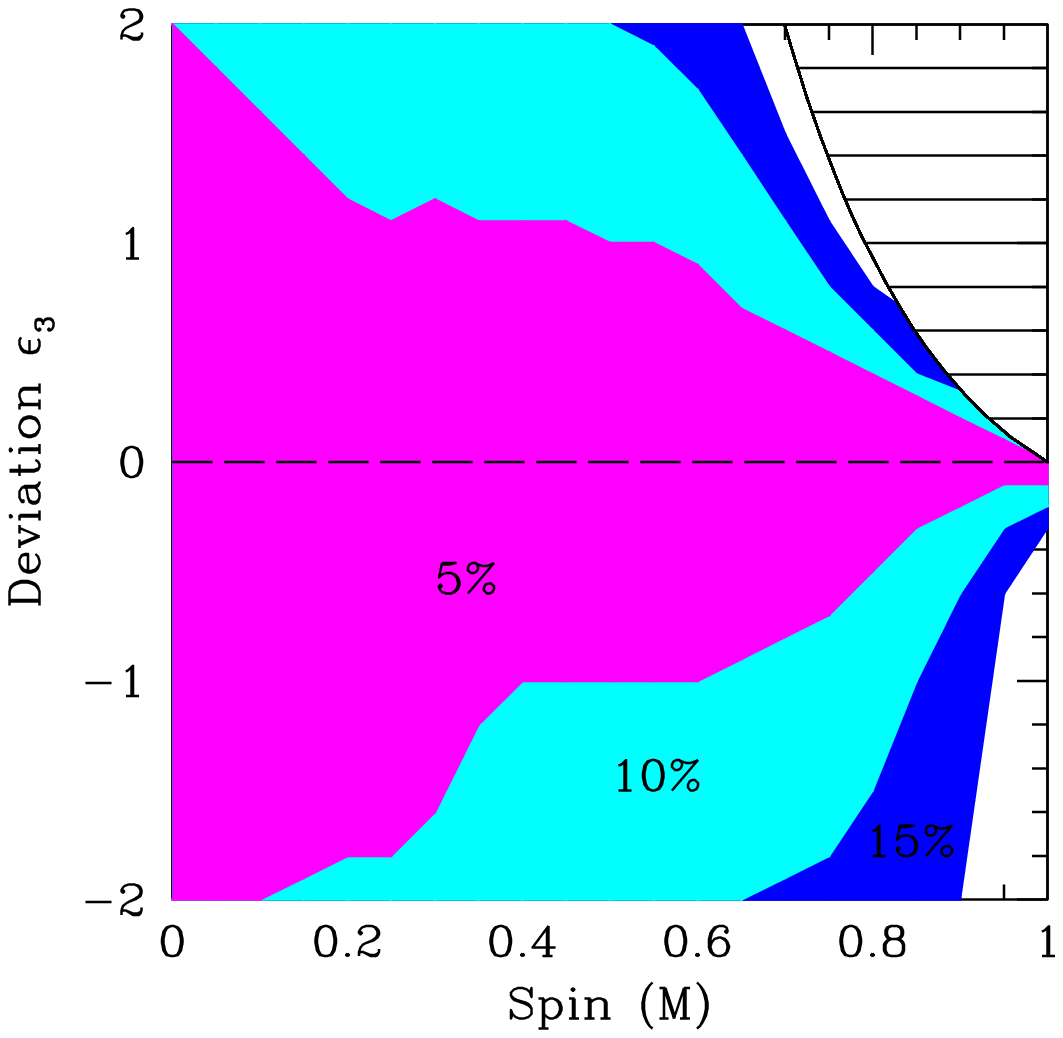,height=3.2in}
\end{center}
\caption{Contours of the precision required in future observations of iron lines to place constraints on deviations from the Kerr metric, for different values of the black-hole spin. Top: $\theta_0=30^\circ$. Bottom: $\theta_0=60^\circ$. The precision requirements are lower at higher spins thanks to the sensitive dependence of the line profiles on the deviation parameter $\epsilon_3$. At both inclinations, the effect of changing the deviation parameter $\epsilon_3$ is similar and constraints $|\epsilon_3| \lesssim 1$ can be obtained at a precision level of $\sim 5\%$ for values of the spin $a\gtrsim0.5M$. The shaded region marks the part of the parameter space, where an ISCO does not exist (see JP11).}
\label{precision}
\end{figure}

In Figure~\ref{precision}, we plot contours of the relative difference $\sigma$ over the ranges $0 \leq a/M \leq 1$ and $-2 \leq \epsilon_3 \leq 2$ for inclination angles $\theta_0=30^\circ$ and $\theta_0=60^\circ$. In both cases, we held the other parameters fixed at values $\alpha=3$ and $r_{\rm out}=100M$.

The relative difference is similar at both inclinations and a precision of $\sim 5\%$ leads to constraints on the parameter $|\epsilon_3|$ to less than order unity at spin values $a\gtrsim0.5M$. At a given level of measured precision constraints on the deviation parameter of the spacetime are significantly more tight at higher spin values.

A measurement of the spin and the deviation parameter from an iron line profile alone may be improved if an independent measurement of these parameters of the same source can be performed. Quasi-periodic variability is a possible second observable, with which an iron line measurement may be combined. In the following, we investigate the prospects of using such QPOs to reduce the correlation between the spin and the parameter $\epsilon_3$. In this paper, we focus as a proof of principle on one particular QPO model, the diskoseismology model (see Wagoner 2008 for a review), where QPOs are identified as so-called gravity modes (g-modes; Perez et al. 1997) and corrugation modes (c-modes; Silbergleit et al. 2001). Similar conclusions can be reached for other QPO models, e.g., Klu\'zniak \& Abramowicz (2001); Abramowicz et al. (2003); Rezzolla et al. (2003). We extend our previous analysis of the diskoseismology model in the context of the quasi-Kerr metric (Johannsen \& Psaltis 2011a) to our new metric, which allows us to discuss the dynamical frequencies of accretion disks for arbitrary values of the spin.

Here, we will only discuss the fundamental g- and c-modes. For any given values of the spin and the parameter $\epsilon_3$, the fundamental g-mode occurs at the radius where the radial epicyclic frequency has a maximum (Perez et al. 1997), while the fundamental c-mode corresponds to the Lense-Thirring frequency evaluated at the ISCO (Silbergleit et al. 2001). We derive expressions for these modes in our metric in the appendix.

\begin{figure}[ht]
\begin{center}
\psfig{figure=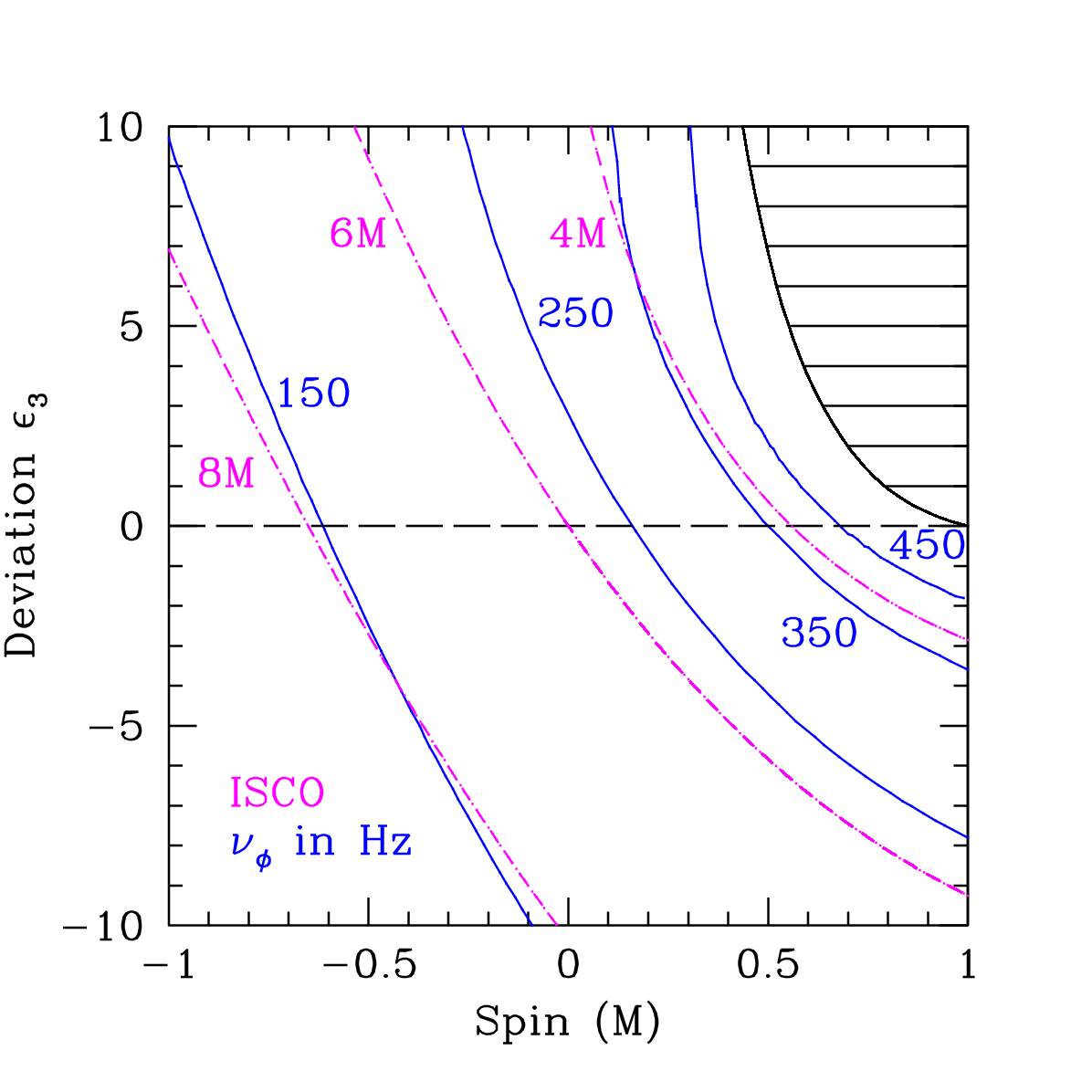,height=3.2in}
\end{center}
\caption{Contours of constant Keplerian frequency evaluated at the ISCO for a 10$M_\odot$ black hole as a function of the spin and the parameter $\epsilon_3$. The Keplerian frequency increases with increasing values of the spin and decreasing values of the parameter $\epsilon_3$. For comparison, we also plot contours of constant ISCO radius. The shaded region marks the part of the parameter space, where an ISCO does not exist (see JP11).}
\label{Keplerian}
\end{figure}

\begin{figure}[ht]
\begin{center}
\psfig{figure=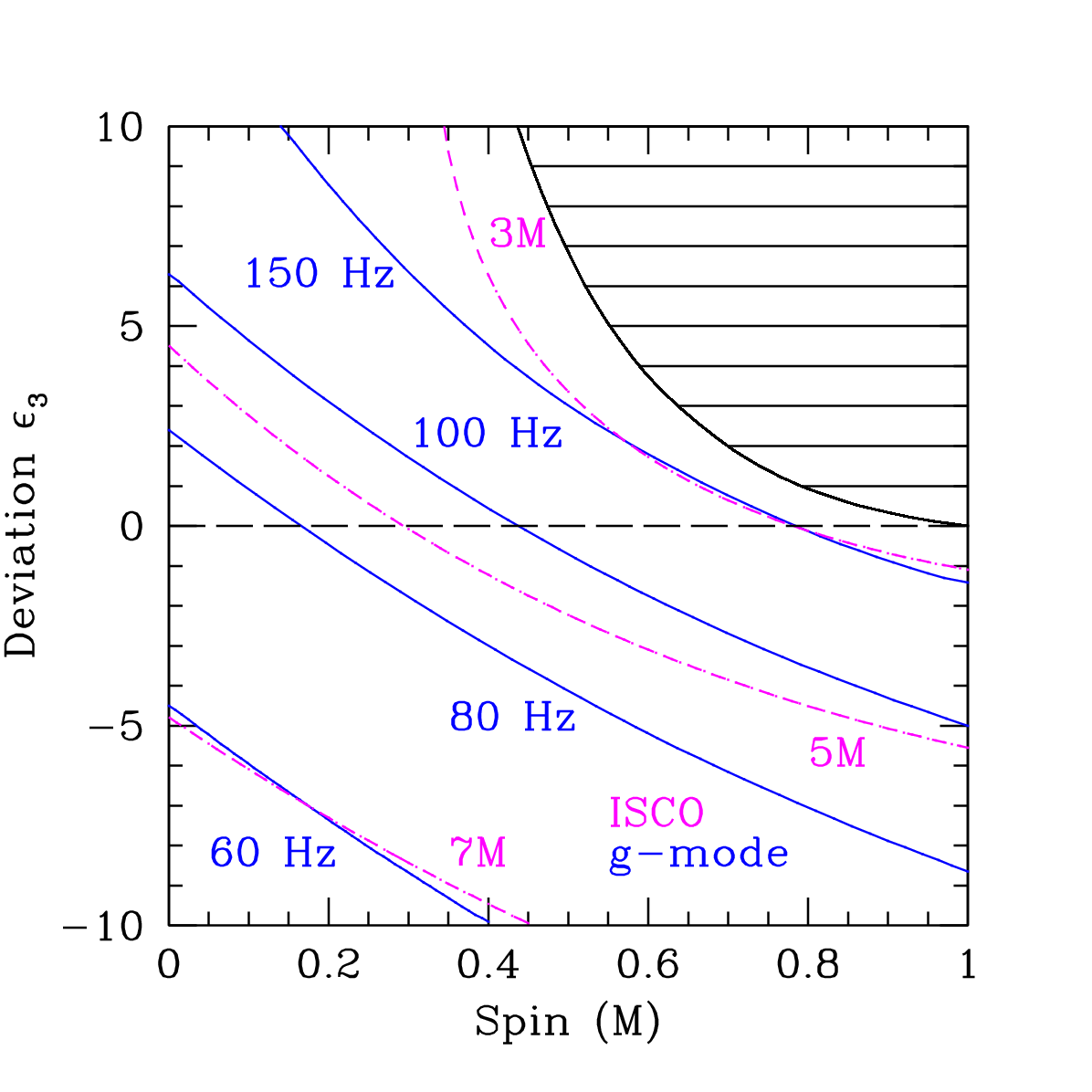,height=3.2in}
\psfig{figure=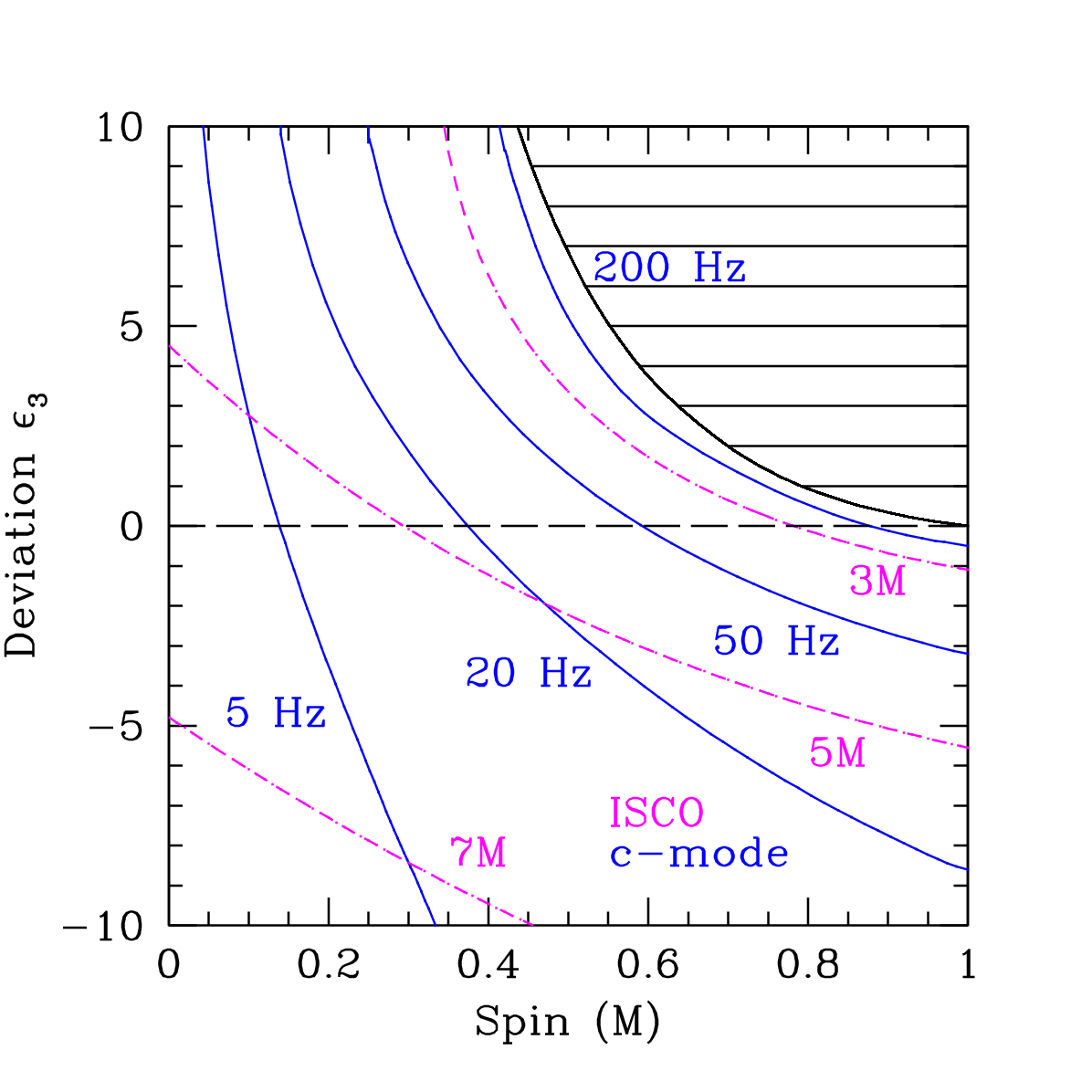,height=3.2in}
\end{center}
\caption{Contours of constant (top) g-mode and (bottom) c-mode frequencies for a 10$M_\odot$ black hole as a function of the spin and the parameter $\epsilon_3$. Both frequencies increase with increasing values of the spin and decreasing values of the parameter $\epsilon_3$. For comparison, we also plot contours of constant ISCO radius. The shaded region marks the part of the parameter space, where an ISCO does not exist (see JP11).}
\label{modes}
\end{figure}

In Figures~\ref{Keplerian} and \ref{modes}, we plot contours of constant Keplerian, g-mode, and c-mode frequencies for a $10M_\odot$ black hole as a function of the spin and the parameter $\epsilon_3$. In all three cases, the contours follow curves of decreasing values of the parameter $\epsilon_3$ for increasing values of the spin. We also plot contours of constant ISCO radius for comparison (see Figure~5 in JP11).

Up to intermediate values of the spin $a\lesssim0.5M$, we recover qualitatively the same dependence of these frequencies on the spin and the deviation parameter as we found for the corresponding frequencies in the quasi-Kerr metric (Johannsen \& Psaltis 2011a). Both the Keplerian frequency and the fundamental g-mode vary as a function of the spin and the parameter $\epsilon_3$ in a manner that is very similar to the dependence of the ISCO radius on these parameters unless the deviation from the Kerr metric is very large. In contrast, the contours of constant c-mode frequency are a lot steeper in the $(a,\epsilon_3)$-plane than the contours of constant ISCO radius. At large spins, all three frequencies follow contours that nearly align with the contours of constant ISCO radius.

As a result, at least within this QPO model, the improvement of a measurement of the spin and the parameter $\epsilon_3$ through the combination of spectral and timing techniques will depend on the mode that is identified. For reasonable deviations from the Kerr metric, only low-frequency c-modes (e.g., modes with frequencies below $\sim 10~{\rm Hz}$ for a 10$M_\odot$ black hole, see Figure~\ref{modes}) can significantly reduce the correlation between the spin and the parameter $\epsilon_3$ from an iron line measurement.

\section{Conclusions}

In this series of papers, we have been investigating a framework for testing the no-hair theorem with observations of black holes in the electromagnetic spectrum using the emission from their accretion flows. Initially, we employed a quasi-Kerr spacetime (Glampedakis \& Babak 2006), which is valid for moderately spinning black holes with spins $a\lesssim0.4M$. We are now extending our framework to include black holes with arbitrary spins using our recently constructed metric (JP11). Since mass and spin are the only free parameters of a Kerr black hole, measuring deviations from the Kerr metric allows us to test the no-hair theorem observationally.

In this paper, we simulated profiles of fluorescent iron lines for values of the spin $0\leq a/M \leq 1$ and the parameter $-2 \leq \epsilon_3 \leq 2$. We showed that deviations from the Kerr metric lead to flux shifts that occur primarily at high energies and in the low-energy tail of the line profiles. We demonstrated that these changes can be significant, especially at high spins. At higher inclination angles, these changes can cause a modification of the flux ratio of the two peaks of the line profile, if a ``red'' peak is apparent in the line shape. We identified this change as a potential signature of a violation of the no-hair theorem, which, in practice, has to be disentangled from the effect of changing the emissivity index $\alpha$ and the outer disk radius $r_{\rm out}$.

We also showed that, as in the case of a Kerr black hole, the location of the ``blue'' edge of the line profile depends exclusively on the disk inclination, at least for small to intermediate values of the inclination angle. In these cases, the disk inclination can be measured independently of the spin and the deviation parameter.

We estimated the required precision of future X-ray missions such as {\em Astro-H} or {\em ATHENA}$+$ in order to be able to test the no-hair theorem with fluorescent iron lines for disk inclinations of $\theta_0=30^\circ$ and $\theta_0=60^\circ$. As an example, we showed that at both inclination angles, measurements with a precision of $\sim 5\%$ can constrain the parameter $|\epsilon_3|$ to order unity for values of the spin $a\gtrsim0.5M$.

However, we also showed that line profiles from spacetimes with different spins and values of the parameter $\epsilon_3$, but with identical locations of the ISCO, are practically indistinguishable from each other. We argued that this correlation can be partially reduced in combination with other observables, which do not depend in the same manner on the location of the ISCO, such as QPOs.

A prime candidate for a test of the no-hair theorem with fluorescent iron lines is the supermassive black hole in the active galaxy 1H0707-495. Recent observations with {\em XMM-Newton} revealed both broad iron K- and L-line transitions (Fabian et al. 2009). Assuming a Kerr black hole, these observations found $a>0.98M$ at an inclination of $\theta_0=55.7^\circ$ (Fabian et al. 2009). Due to its high spin, we expect relatively tight constraints on potential deviations from the Kerr metric from this source.

Two additional promising approaches for testing the no-hair theorem lie in observations of fluorescent iron line profiles in the time domain. First, localized regions of iron line emission follow nearly Keplerian orbits around the black hole and appear as arcs in the dynamical energy spectra (Brenneman et al. 2009). Second, iron line variability due to X-ray reverberation (e.g., Zoghbi et al. 2012) is a measure of the light travel time delay between the corona and the accretion disk. Upcoming X-ray timing missions, such as the {\em Large Observatory For x-ray Timing (LOFT)}, will for the first time have the sensitivity to perform time-resolved observations of iron lines, which can trace these hot spots as they orbit around the black hole as well as measure reverberation time delays with high precision. We will analyze the prospects of these techniques for tests of the no-hair theorem in a future paper.

We thank the referee, C. Reynolds, as well as L. Brenneman for useful comments. This work was supported at the University of Arizona by the NSF CAREER award NSF 0746549. TJ was also supported by a CITA National Fellowship at the University of Waterloo. Research at Perimeter Institute is supported by the Government of Canada through Industry Canada and by the Province of Ontario through the Ministry of Research and Innovation.

\appendix
\section{Properties of a Kerr-like Black Hole Spacetime}

In this appendix, we briefly summarize the explicit form and some of the properties of our metric (JP11) and we derive the Keplerian and epicyclic frequencies of a particle on a circular equatorial orbit. In Boyer-Lindquist-like coordinates, the metric can be expressed by the line element
\ba
ds^2 = && -[1+h(r,\theta)] \left(1-\frac{2Mr}{\Sigma}\right)dt^2 -\frac{ 4aMr\sin^2\theta }{ \Sigma }[1+h(r,\theta)]dtd\phi + \frac{ \Sigma[1+h(r,\theta)] }{ \Delta + a^2\sin^2\theta h(r,\theta) }dr^2 + \Sigma d\theta^2 \nonumber \\
&& + \left[ \sin^2\theta \left( r^2 + a^2 + \frac{ 2a^2 Mr\sin^2\theta }{\Sigma} \right) + h(r,\theta) \frac{a^2(\Sigma + 2Mr)\sin^4\theta }{\Sigma} \right] d\phi^2,
\label{metric}
\ea
where
\ba
\Sigma &\equiv& r^2 + a^2 \cos^2\theta, \\
\Delta &\equiv& r^2 - 2Mr + a^2, \\
h(r,\theta) &\equiv& \sum_{k=1}^\infty \left( \epsilon_{2k} + \epsilon_{2k+1}\frac{Mr}{\Sigma} \right) \left( \frac{M^2}{\Sigma} \right)^{k}
\label{h(r,theta)}
\ea
with the free parameters $\epsilon_k$. We will use this metric with only one additional parameter $\epsilon_3$, so that the function $h(r,\theta)$ reduces to
\be
h(r,\theta) = \epsilon_3 \frac{M^3 r}{\Sigma^2}.
\label{hchoice}
\ee

For a particle on a circular equatorial orbit, its energy and axial angular momentum in units of the rest mass $\mu$ of the particle are given by the expressions (JP11; Johannsen 2013b)
\be
\frac{E}{\mu} = \sigma_1 \frac{1}{r^6}\sqrt{ \frac{P_1 + P_2}{P_3} },
\label{energy}
\ee
\be
\frac{L_z}{\mu} = \frac{1}{r^4 P_6} \bigg[ \sigma_2 \sqrt{\frac{P_5}{P_3} } + \sigma_3 6a M (r^3 + \epsilon_3 M^3) \sqrt{ \frac{P_1 + P_2}{P_3} } \bigg],
\label{angularmomentum}
\ee
where $\sigma_{1-3}\equiv\pm1$. The sign of the parameter $\sigma_1$ depends on the value of the deviation parameter, while the signs of the parameters $\sigma_2$ and $\sigma_3$ depend on the root structure of the functions $P_1$ to $P_6$, which are given below. In most cases, $\sigma_1=+1$, $\sigma_2=\pm1$ and $\sigma_3=\mp1$, where the upper signs refer to prograde orbits, while the lower signs refer to retrograde orbits. These parameters can only change sign at radii that are very close to the ISCO radius. In particular, the parameter $\sigma_1$ can only change sign for values of the parameter $\epsilon_3\geq\epsilon_3^{\rm bound}$, where the bound on the parameter $\epsilon_3$ is given by the expression (Johannsen 2013a)
\ba
\epsilon_3^{\rm bound}(a) &=& \frac{1}{3125(a/M)^2} \bigg[ 1024 \left( 4 + \sqrt{16-15(a/M)^2} \right) - 160(a/M)^2 \left(40+7\sqrt{16-15(a/M)^2} \right) \nonumber \\
&& + 150 (a/M)^4 \left(15+\sqrt{16-15(a/M)^2} \right) \bigg].
\label{eq:bound}
\ea
See Johannsen (2013b) for a full discussion of the energy $E$, the axial angular momentum $L_z$ and the roots of the functions $P_1$ to $P_6$. The bound given by Equation~(\ref{eq:bound}) also determines the region of the parameter space, where an ISCO exists. An ISCO exists everywhere except for values of the parameter $\epsilon_3\geq\epsilon_3^{\rm bound}$ if $a>0$ (JP11; Johannsen 2013a).

The functions $P_1$ to $P_6$ are given by the following expressions (Johannsen 2013b):
\ba
P_1 = &&a^2 M r^4 \left(\epsilon_3 M^3 +r^3\right)^2 \bigg\{12\epsilon_3 a^2 M^3 \left(\epsilon_3 M^3+r^3\right)^2 \nonumber \\
&&-r^4 \left[2\epsilon_3 M^2 r^3 \left(3 r^2-8 M^2\right)+\epsilon_3^2 M^5 \left(40 M^2-48 M r+15 r^2\right)+4 r^6 (3 r-5 M)\right]\bigg\}, \\
P_2 = &&2 \bigg\{2 r^4\left(r^{20} \mp M P_4\right)+M r^{12} \bigg\{ 2 r^9 \left(-12 M^2+16 M r-7r^2\right) \nonumber \\
&&+\epsilon_3 M^2 (r-2 M)^2 \left[\epsilon_3^2 M^6 (5r-12 M)-6\epsilon_3 M^3 r^3 (5 M-2 r)-3 r^6 (8 M-3r)\right]\bigg\}\bigg\}, \\
P_3 = &&r^4 \left(12\epsilon_3 M^4 -5\epsilon_3 M^3 r +6 M r^3-2 r^4\right)^2-8 a^2 M \left(\epsilon_3 M^3 +r^3\right)^2 \left(5\epsilon_3 M^3 +2 r^3\right), \\
P_4 = &&
\left\{\begin{array}{l}
-a B~~~{\rm if~}n_{41}<r<n_{42}\\
a B~~~~~{\rm else}
\end{array}\right.
, \\
P_5 = && M \left(r^3+\epsilon_3 M^3\right)^2 \bigg\{12\epsilon_3 a^6 M^3 \left(\epsilon_3 M^3 -2 r^3\right)^2 \left(\epsilon_3 M^3+r^3\right)^4+a^4 r^4 \left(\epsilon_3 M^3+r^3\right)^2 \nonumber \\
&&\left(-40\epsilon_3^4 M^{13}+40\epsilon_3^4 M^{12} r -15\epsilon_3^4 M^{11} r^2+128\epsilon_3^3 M^{10} r^3 -296\epsilon_3^3 M^9 r^4 +54\epsilon_3^3 M^8 r^5-924\epsilon_3^2 M^7 r^6 \right.\nonumber \\
&&\left.+276\epsilon_3^2 M^6 r^7 -36\epsilon_3^2 M^5 r^8 -880\epsilon_3 M^4 r^9 +304\epsilon_3 M^3 r^{10}-24\epsilon_3 M^2 r^{11}-112 M r^{12}+16 r^{13}\right) \nonumber \\
&&-2 a^2 r^8 \big[48 \epsilon_3^5 M^{17} -12\epsilon_3^5 M^{16} r -52\epsilon_3^5 M^{15} r^2+3\epsilon_3^4 M^{14} r^3 (5\epsilon_3+88)-720\epsilon_3^4 M^{13} r^4 +298\epsilon_3^4 M^{12} r^5 \nonumber \\
&& -3\epsilon_3^3 M^{11} r^6 (13 \epsilon_3+480)+516\epsilon_3^3 M^{10} r^7 +2\epsilon_3^3 M^9 r^8 -6\epsilon_3^2 M^8 r^9 (3\epsilon_3+508)+2292\epsilon_3^2 M^7 r^{10} \nonumber \\
&& -628 \epsilon_3^2 M^6 r^{11}+12\epsilon_3 M^5 r^{12} (5\epsilon_3-134)+1188\epsilon_3 M^4 r^{13} -296\epsilon_3 M^3 r^{14} +24 M^2 r^{15} (\epsilon_3-9)+120 M r^{16} \nonumber \\
&&-16r^{17}\big]-r^{14} \left(\epsilon_3 M^3 +6 M r^2-2 r^3\right)^2 \nonumber \\
&&\left(96\epsilon_3^2 M^7 -76\epsilon_3^2 M^6 r +15\epsilon_3^2 M^5 r^2+72\epsilon_3 M^4 r^3 -44\epsilon_3 M^3 r^4 +6\epsilon_3 M^2 r^5 +12 M r^6-4 r^7\right)\bigg\} \nonumber \\
&& -4 P_4 \left[ a^2 \left(\epsilon_3 M^3 -2 r^3\right)^2 \left(\epsilon_3 M^3+r^3\right)+6\epsilon_3 M^3 r^5 \left(\epsilon_3 M^3 +6 M r^2-2 r^3\right)\right], \\
P_6 = && -\epsilon_3^2 M^6 -6\epsilon_3 M^4 r^2 +\epsilon_3 M^3 r^3 -6 M r^5+2 r^6,
\ea
where
\be
B \equiv \sqrt{M \left(\epsilon_3 M^3 +r^3\right)^6 \left(9\epsilon_3^2 a^2 M^5+16\epsilon_3 M^3 r^4 -6\epsilon_3 M^2 r^5 +4 r^7\right) \left[a^2 \left(\epsilon_3 M^3 +r^3\right)+r^4 (r-2 M)\right]^2}
\ee
and where $n_{41}$ and $n_{42}$ denote the smaller and greater (real) root of the function $P_4$, respectively, if real roots exist for $r>0$.

Next, we derive the Keplerian orbital frequency of a particle with energy $E$ and axial angular momentum $L_z$. A particle with a rest mass $\mu$ has the 4-momentum
\be
p^\alpha = \mu \frac{dx^\alpha}{d\tau},
\ee
where $E=-p_t$ and $L_z=p_\phi$. Hereafter, we will set the rest mass equal to unity. Using the conservation of the energy and axial angular momentum, we solve for the momentum components $p^t$ and $p^\phi$ and obtain
\ba
p^t = - \frac{ g_{\phi\phi}E + g_{t\phi}L_z }{ g_{tt}g_{\phi\phi} - g_{t\phi}^2 }, \\
p^\phi = \frac{ g_{t\phi}E + g_{tt}L_z }{ g_{tt}g_{\phi\phi} - g_{t\phi}^2 }.
\label{velocity}
\ea
For the Keplerian frequency, we derive the expression
\be
\Omega_\phi \equiv \frac{d\phi}{dt} = \frac{p^\phi}{p^t} = - \frac{g_{t\phi}E + g_{tt}L_z}{g_{\phi\phi}E + g_{t\phi}L_z}.
\ee
This expression is lengthy, and we do not write it here explicitly. In the case $\epsilon_3=0$, the expression of the Keplerian frequency reduces to the familiar Kerr Keplerian frequency
\be
\Omega_\phi^K = \pm \frac{ \sqrt{M} }{ r^{3/2} \pm a \sqrt{M} }.
\ee

Finally, we compute the radial and vertical epicyclic frequencies as well as the Lense-Thirring frequency. Our derivation is analogous to the one in Johannsen \& Psaltis (2011a). We use the conservation of the particle's rest mass to define an effective potential
\be
V_{\rm eff} \equiv \frac{1}{2}\left[ -g_{\rm tt}(p^{\rm t})^2 - 2g_{\rm t\phi}p^{\rm t}p^{\rm \phi} - g_{\rm \phi\phi}(p^{\rm \phi})^2 - 1 \right].
\ee

Radial and vertical motions around a circular equatorial orbit at a radius $r_0$ are governed by the equations
\ba
\frac{1}{2} \left( \frac{dr}{dt} \right)^2 &=& \frac{ V_{\rm eff} }{ g_{\rm rr}(p^{\rm t})^2 } \equiv V_{\rm eff}^{\rm r}, \label{Veffr} \\
\frac{1}{2} \left( \frac{d\theta}{dt} \right)^2 &=& \frac{ V_{\rm eff} }{ g_{\rm \theta\theta}(p^{\rm t})^2 } \equiv V_{\rm eff}^{\rm \theta},
\label{Vefftheta}
\ea
respectively.

We, then, introduce small perturbations $\delta r$ and $\delta \theta$ and take the coordinate time derivative of Equations (\ref{Veffr}) and (\ref{Vefftheta}), which yields
\ba
\frac{d^2(\delta r)}{dt^2} &=& \frac{ d^2 V_{\rm eff}^{\rm r} }{ dr^2 } \delta r, \\
\frac{d^2(\delta \theta)}{dt^2} &=& \frac{ d^2 V_{\rm eff}^{\rm \theta} }{ d\theta^2 } \delta \theta.
\ea
From these expressions, we derive the radial and vertical epicyclic frequencies as
\ba
\kappa_{\rm r}^2 &=& -\frac{ d^2 V_{\rm eff}^{\rm r} }{dr^2}, \\
\label{kappa}
\Omega_{\rm \theta}^2 &=& -\frac{ d^2 V_{\rm eff}^{\rm \theta} }{ d\theta^2 },
\label{omegatheta}
\ea
where the second derivatives are evaluated at $r=r_0$. These expressions are lengthy, and we do not write them here explicitly.

Having the expressions of the Keplerian and vertical epicyclic frequencies at hand, the Lense-Thirring frequency is given by
\be
\Omega_{\rm LT} \equiv \Omega_\phi - \Omega_\theta.
\label{LenseThirring}
\ee

\end{document}